\documentclass[twocolumn,trackchanges]{aastex701}

\usepackage{CJKutf8}
\usepackage{microtype}
\usepackage[T1]{fontenc}
\usepackage{amsmath}
\usepackage{multirow}
\usepackage{bm}
\usepackage{dsfont}
\usepackage{mathrsfs}
\usepackage{gensymb}
\usepackage[dvipsnames]{xcolor}
\usepackage{soul}
\usepackage{fontawesome}
\usepackage{graphicx}
\usepackage{booktabs}
\usepackage{placeins}
\usepackage[figure, figure*]{hypcap}
\usepackage{todonotes}
  \setuptodonotes{
    backgroundcolor=lightgray!50,
    linecolor=lightgray,
    bordercolor=none,
    tickmarkheight=0.2em,
  }

\newcommand{\gkai}[1]{\begin{CJK*}{UTF8}{gkai}\raisebox{.1em}{(}#1\raisebox{.1em}{)}\end{CJK*}}

\bmdefine{\vx}{x}
\bmdefine{\vk}{k}

\newcommand{\pard}[2]{\frac{\partial #1}{\partial #2}}

\newcommand{\Dfour}{$\mathrm{D}_4$}
\newcommand{\anacal}{\texttt{AnaCal}}
\newcommand{\xlens}{\texttt{xlens}}

\shorttitle{D$_4$CNN$\times$AnaCal}
\shortauthors{Lin et al.}

\begin{document}

\title{D$_4$CNN$\times$AnaCal: \\ Physics-Informed Machine Learning for Accurate and Precise Weak Lensing Shear Estimation}

\correspondingauthor{Shurui Lin}
\email{shuruil3@illinois.edu}

\author[orcid=0009-0000-5381-7039]{Shurui Lin\gkai{林书睿}}
\affiliation{Department of Astronomy, University of Illinois at
Urbana-Champaign, 1002 West Green Street, Urbana, IL 61801, USA}
\email[]{shuruil3@illinois.edu}
\author[orcid=0000-0003-2880-5102,gname=Xiangchong, sname='Li']{Xiangchong Li}
\affiliation{Brookhaven National Laboratory, Bldg 510, Upton, New York 11973, USA}
\email[]{xli6@bnl.gov}
\author[]{Ji Li}
\affiliation{Department of Computer Science, University of Illinois at Urbana-Champaign, 201 North Goodwin Avenue, Urbana, IL 61801, USA}
\email[]{}
\author[]{Shengcao Cao}
\affiliation{Department of Computer Science, University of Illinois at Urbana-Champaign, 201 North Goodwin Avenue, Urbana, IL 61801, USA}
\email[]{}
\author[orcid=0000-0003-0049-5210]{Xin Liu}
\affiliation{Department of Astronomy, University of Illinois at Urbana-Champaign, 1002 West Green Street, Urbana, IL 61801, USA}
\affiliation{National Center for Supercomputing Applications, University of Illinois at Urbana-Champaign, Urbana, IL 61801, USA}
\affiliation{Center for Artificial Intelligence Innovation, University of Illinois at Urbana-Champaign, 1205 West Clark Street, Urbana, IL 61801, USA}
\email[]{xinliuxl@illinois.edu}
\author[]{Yu-Xiong Wang}
\affiliation{Department of Computer Science, University of Illinois at Urbana-Champaign, 201 North Goodwin Avenue, Urbana, IL 61801, USA}
\email[]{}

\begin{abstract}
Traditional weak gravitational lensing shear estimators are carefully calibrated but struggle to fully capture realistic galaxy morphologies, point-spread-function (PSF) effects, blending, and noise in deep surveys, while blindly trained machine learning (ML) models can introduce significant calibration biases. 
Here we construct a fully D$_4$-equivariant deep neural network for galaxy shape measurement whose architecture enforces symmetry under 90$^{\circ}$ rotations and mirror transformations, and adopt the Analytical Calibration framework (AnaCal) to calibrate the model using its backpropagated gradients. 
For isolated galaxies in LSST-like single-band simulations, we demonstrate that our approach achieves $\sim$10\% lower shape noise than the traditional moment-based Fourier Power Function Shapelets estimator in the high-noise regime, equivalent to a 23\% gain in effective galaxy number density, while simultaneously achieving multiplicative biases consistent with zero across a wide range of noise levels, PSF sizes and ellipticities, and magnitude selection cuts, with all measurements satisfying $|m| {<} 10^{-3}$ (i.e., within the 0.2\% LSST requirement) and most at the ${\sim}10^{-4}$ level. 
We demonstrate this framework on isolated single-band galaxy images with Gaussian noise and known PSF, establishing a rigorous, physics-informed foundation for future extensions of ML-based shear estimation to blended sources and multi-band observations in Stage-IV surveys. 
All codes and data products will be made publicly available upon acceptance.

\end{abstract}

\keywords{
    \uat{Cosmology}{343} ---
    \uat{Weak gravitational lensing}{1797} ---
    \uat{Astronomy image processing}{2306} ---
    \uat{Convolutional neural networks}{1938}
}

\section{Introduction}
\label{sec:intro}

Weak gravitational lensing by large-scale structure (i.e., cosmic shear) has emerged as a premier probe of dark energy and cosmic structure growth \citep{Mellier1999,Bartelmann2001,Kilbinger2015}, but its potential can only be realized with exquisitely accurate shear estimation \citep{Mandelbaum2018}.
Aside from the significant statistical noise (``shape noise'') induced by the dispersion of intrinsic galaxy shapes,
central challenges in shear estimation come from pixel noise, galaxy selection, and point-spread function (PSF), each of which could potentially cause systematic biases.
Achieving the sub-percent accuracy required by Stage-IV surveys (Euclid \citep{EuclidCollaboration2024c}; Rubin/LSST \citep{Ivezic2019}; and Roman \citep{Spergel2015}) demands careful control of these biases in galaxy shear estimation.
This directly demands that shape measurement methods must be calibrated for the various biases mentioned above.
However, the calibration step might be resource-intensive and can itself introduce uncertainties \citep{DESY1Shape2018},
motivating the development of self-calibration techniques and fast shape measurement methods.

Over the past decade, several innovative self-calibration approaches have been proposed.
One class of methods is \texttt{METACALIBRATION} \citep{Huff2017, Sheldon2017}, which applies known shear distortions to galaxy images to directly compute the response of shear estimators.
By measuring each galaxy's change in measured ellipticity under artificially sheared versions of the image, \texttt{METACALIBRATION} calibrates biases using the data itself.
An extension of this idea is the \texttt{METADETECTION} technique \citep{Sheldon2020}, which incorporates object detection into the \texttt{METACALIBRATION} process to account for detection biases by rerunning the detection and measurement on sheared images.
These methods provide a form of numerical self-calibration that are being implemented for Stage-IV surveys \citep{Yamamoto2023,Zhang2023c}.
An important advance beyond these numerical self-calibration methods is the Analytic Calibration method (\anacal{}) \citep{LiMandelbaum2023,Li2024}.
\anacal{} enables a fully analytical calculation of shear response, propagating the derivatives of each pixel's intensity through the detection and measurement process to evaluate the final ellipticity measurement's response to a small shear distortion.
This approach can account for major sources of bias (detection, selection, galaxy model, and PSF anisotropy) and achieve unbiased shear estimates to second order in shear without requiring large suites of calibrated simulations.
\citet{LiMandelbaum2023} has demonstrated the accuracy and efficiency of \anacal{} for the Hyper Suprime-Cam Survey \citep{Aihara2018a} with realistic image simulations.
Furthermore, \citet{Li2025} has presented an improved formulation of the \anacal{} method targeting LSST requirements on simulations, incorporating a rigorous correction for noise bias, making the analytical calibration framework even more robust and up-to-date.

In parallel, machine learning (ML) techniques have emerged as a promising tool for weak lensing shear estimation.
First, they can leverage subtle, higher-order morphological information in galaxy images--features beyond simple moments--that could help reduce shape noise.
Second, ML methods can adapt flexibly to complex observational conditions, including non-Gaussian noise, PSF variability, and source blending, which increase the uncertainty for analytical estimators.
Third, with the two above advantages, they are computationally efficient, scaling favorably to the billions of sources expected from Euclid, LSST, and Roman.

Previous ML studies showed that supervised ML algorithms can learn to extract shape information directly from galaxy images, potentially surpassing some limitations of classical methods.
\citet{Tewes2019} demonstrated that a simple feed-forward network can learn to compensate for noise and PSF effects, achieving percent-level accuracy on GREAT3 challenge images \citep{GREAT3}.
Subsequent studies developed more advanced pipelines.
\citet{Ribli2019,Ribli2019a} applied convolutional neural networks (CNNs) to predict ellipticities with performance comparable to standard methods, while \citet{Pujol2019} and follow-up work \citep{Pujol2020,Pujol2020a} used shallow ANNs and showed that ML-based measurements can be calibrated to high accuracy.
Building on these, \citet{Springer2020} reduced the impact of shape noise by training an ensemble of ANNs on large numbers of galaxies, effectively averaging over intrinsic shape variations to push beyond the traditional shape-noise limit.
\citet{Zhang2024} presented \textsc{FORKLENS}, a two-stage deep learning framework that first regresses shear using a CNN and then applies a learned calibration network to correct residual biases.
These studies indicate that modern neural networks, if properly trained and calibrated, can rapidly measure shear with low bias and scatter.

Despite their promise, achieving the rigorous standards of Stage-IV surveys remains a significant challenge.
For instance, the LSST Dark Energy Science Collaboration (DESC) mandates that shear estimation biases be controlled consistently at the subpercent level \citep{LSSTSRD}, a benchmark that current ML-based estimators are still striving to meet with the necessary stability.
The critical reason is that many of these frameworks treat shear estimation as a general image-regression task, effectively neglecting two fundamental physical priors of weak lensing shear \citep{LiMandelbaum2023}.
First, they often fail to encode the symmetry of shear--e.g., the mathematical requirement that the signal remains invariant under a $180^\circ$ rotation.
Second, they rarely exploit the fact that shear is a perturbative signal ($\gamma \ll 1$) and the calibration relies on a continuous gradient of the shape measurement.
Due to ignorance of these physical constraints, existing estimators usually rely on specific weights with strong dataset dependence and fail to generalize when the observed data deviates even slightly from simulations.

To address these limitations, we introduce a physics-informed machine learning framework \citep{Karniadakis2021} that moves beyond ``black-box'' regression by embedding the fundamental principles of weak lensing shear directly into the model.
Specifically, we design an equivariant CNN architecture (hereafter D$_4$CNN) whose predicted ellipticities transform exactly as expected under $90^{\circ}$ rotations and mirror flips of the galaxy image, eliminating the non-symmetric contributions to the output.
This built-in equivariance strongly suppresses second-order bias terms, making the ellipticity–shear mapping $e(\gamma)$ highly linear.
In addition, we apply the Analytic Calibration approach to our network's predictions. With the smoothly differentiable nature of our model, we can analytically compute the shear response (Jacobian) of the output with respect to a shear distortion of the input, analogous to the \anacal{} technique.
Using this computed response matrix, we calibrate the raw network predictions to correct any first-order bias. 
The combination of \Dfour{} symmetry and analytical calibration yields a shear estimator that is unbiased to the second order in shear without requiring large simulation-based corrections. 
To our knowledge, this is the first application of a fully \Dfour{}-equivariant deep neural network (i.e., one whose architecture hard-codes the spin-2 symmetry of galaxy ellipticity) to weak lensing shear estimation, and the first to pair such a network with an analytical differentiable self-calibration scheme. 
While equivariant networks \citep{Cohen2016} have been applied to various astronomy tasks \citep[e.g.,][]{Dieleman2015,Scaife2021,Pandya2023,Dai2022,Perraudin2019,Tripathi2025}, the combination of hard-coded \Dfour{} equivariance with AnaCal-style gradient-based calibration for unbiased shear recovery is novel.
As a first demonstration of this framework, we focus on isolated galaxies in single-band imaging with Gaussian noise and a known PSF (conditions under which the theoretical guarantees of our calibration scheme can be most cleanly validated). Extensions to blended sources, chromatic PSFs, and multi-band data are natural next steps and are discussed in \autoref{sec:conclusion}.

With realistic simulated galaxy images, we show that the network achieves shear recovery with negligible bias (well below $0.2\%$ to meet Stage-IV survey requirements) and improved shape noise. This approach thus combines the strengths of machine learning--flexibility and high signal-to-noise efficiency--with the rigorous bias control of analytical self-calibration methods.

The rest of the paper is organized as follows.
\autoref{sec:ML_Anacal} reviews the shear response formalism and describes how analytical calibration can be applied to ML models.
\autoref{sec:Data} shows the details of the image simulation and calibration method.
\autoref{sec:ML} introduces the equivariant CNN architecture and training procedure.
In \autoref{sec:Result} and \autoref{sec:shape_noise}, we evaluate the performance of our method on simulated data, including tests of bias and uncertainty.
We further investigate the impact of equivariance and gradient smoothness in \autoref{sec:sym&grad}.
Finally, \autoref{sec:conclusion} discusses the implications of our results, including the path toward blended sources and scene-native calibration, and situates \textsc{D$_4$CNN}$\times$\textsc{AnaCal} within the broader landscape of shear pipelines for Euclid, LSST, and Roman.

\section{Methods}

\subsection{Shear Calibration for ML model}
\label{sec:ML_Anacal}
In this section, we quickly review the analytical form of shear response following \citet{LiMandelbaum2023} and \citet{Li2025}. With this, we further explore how this technique can be implemented in machine learning models, together with the required symmetry of the machine learning model.

\subsubsection{Observables and Shear Responses}
\label{subsec:shear_rsp}
We start with a set of linear observables that are linear combinations of pixel values to get galaxy ellipticity, following \cite{Li2025}.
We define the observed image smeared by PSF as $f_p(\mathbf{x})$, where $p(\mathbf{x})$ is the corresponding PSF model.
And the Fourier transformed image and PSF are:
\begin{align}
f_p(\mathbf{k}) &= \iint \mathrm{d}^2x \, f_p(\mathbf{x})
   \exp(-i \, \mathbf{k} \cdot \mathbf{x}) , \\
p(\mathbf{k}) &= \iint \mathrm{d}^2x \, p(\mathbf{x})
   \exp(-i \, \mathbf{k} \cdot \mathbf{x}) .
\end{align}
To remove anisotropy from the PSF and noise and calibrate noise bias, we use re-smoothed, re-noised image, which is reconvolved with a Gaussian PSF kernel and added additional noise to eliminate noise bias.
In this case, the most common linear observable input for the model is the re-smoothed image $f_h(\mathbf{x})$:
\begin{align}
    f_h(\mathbf{x}) = \mathcal{F}^{-1}
    \left[
    \left(
    \frac{f_p(\vk)}{p(\vk)}-\frac{n(\vk)}{p^{90}(\vk)}
    \right)*h(\vk)
    \right],
\label{eq:resmooth_img}
\end{align}
where $h(\vk) = \mathrm{exp}(-k^2\sigma_h^2/2)$ is the Gaussian filter with a standard deviation $\sigma_h$. $n(\vk)$ is the noise image in Fourier space and $p^{90}(\vk)$ is the PSF after 90-degree rotation \citep{LiMandelbaum2025}.
Under this notation, the machine learning shape measurement model can be expressed as a function $M\left(f_h(\vx)\right)$ that maps an input re-smoothed image into ellipticity $e_{1,2}$:
\begin{align}
    e_{1,2} = M(f_h(\vx)).
\end{align}
By definition, the ellipticity should be a spin-2 vector under rotation, which would reduce to \Dfour{} (i.e., the mathematical group including 90-degree rotations and mirror) equivariance under the pixelization of the image.
However, this is not generally guaranteed by machine learning models, so the output shape would have a non-symmetric noise term:
\begin{align}
    e_{1,2} = e_\mathrm{D_4}+\delta e_\mathrm{non-D_4},
\end{align}
where $e_\mathrm{D_4}$ stands for the \Dfour{} symmetric part and $\delta e_\mathrm{non-D_4}$ is the non-symmetric noise.

Considering the shear response of the ellipticity from the model, assuming the shape is continuously differentiable to at least the third order:
\begin{align}
e_{1,2}(\gamma_i)
&= e_{1,2}
+ \frac{\partial e_{1,2}}{\partial \gamma_i} \gamma_i
+ \frac{1}{2} \frac{\partial^2 e_{1,2}}{\partial \gamma_i \, \partial \gamma_j} \, \gamma_i \gamma_j \notag \\
&\quad + \mathcal{O}(\gamma^3),
\end{align}
where $\mathbf{\gamma}$ is the shear distortion. The zeroth-order term is the
intrinsic shape. As shown in \cite{Li2024}, \Dfour{} symmetry of ellipticity
would force the expectation of all even-order terms to be zero. As a result,
shear response would become:
\begin{equation}
    \begin{aligned}
    \mathbb{E}\big[\delta e_{1,2}(\gamma_i)\big]
    &= \mathbb{E}[e_\mathrm{non\text{-}D_4}]
    + \frac{\partial}{\partial \gamma_i}\,\mathbb{E}[e_{1,2}] \, \gamma_i \\
    &\quad + \frac{1}{2}\,
    \frac{\partial^2}{\partial \gamma_i \, \partial \gamma_j}
    \,\mathbb{E}\!\left[\delta e_\mathrm{non\text{-}D_4}\right]
    \, \gamma_i \gamma_j \\
    &\quad + \mathcal{O}(\gamma^3) .
\end{aligned}
\end{equation}

So, the second-order non-linearity would completely come from the non-\Dfour{} noise.
Based on this result, as we will show in \autoref{sec:ML}, 
we hard-code the symmetry into the architecture and successfully reduce the non-\Dfour{} noise to the $10^{-6}$ level 
(measured by summing the shape measurements of a random galaxy and its 90°-rotated counterpart), 
limited by the accumulation of float32 arithmetic errors across the eight forward passes rather than by physical asymmetry.
In this case, we can safely ignore the second-order term and keep the leading non-linearity to the third-order:
\begin{align}
\mathbb{E}[e_{1,2}(\gamma_i)]
&= \frac{\partial \mathbb{E}\left[e_{1,2}\right]}{\partial \gamma_i} \, \gamma_i
+ \mathbb{E}[\mathcal{O}(\gamma^3)].
\label{eq:shear_r}
\end{align}

In real observation, we would use the ensemble average of ellipticity $\langle e_{1,2} \rangle$ to estimate the expectation of shape.
With a large enough ensemble that shares the same shear within, $\langle e_{1,2} \rangle$ would serve as an estimator of shear:
\begin{align}
    \langle e_{1,2} \rangle = \frac{\partial \langle e_{1,2} \rangle}{\partial \gamma_i} \gamma_i +\langle \mathcal{O}(\gamma^3) \rangle.
\end{align}
Ignoring the high-order term, we would have an unbiased estimator:
\begin{align}
    \hat\gamma_i = \langle e_{1,2} \rangle \Big/ \frac{\partial \langle e_{1,2} \rangle}{\partial \gamma_i}.
\label{eq:unbiased_esti}
\end{align}
Based on \autoref{eq:unbiased_esti}, we would first discuss the case without selection in \autoref{subsec:anacal}, where we can exchange average and gradient to get the following estimator:
\begin{align}
    \hat\gamma_i = \langle e_{1,2} \rangle \Big/ \left\langle{\frac{\partial e_{1,2}}{\partial \gamma_i}}\right \rangle.
\end{align}

\subsubsection{Analytical Calibration for ML model}
\label{subsec:anacal}

\begin{figure*}[tb]
\centering
\includegraphics[width=1.0\textwidth]{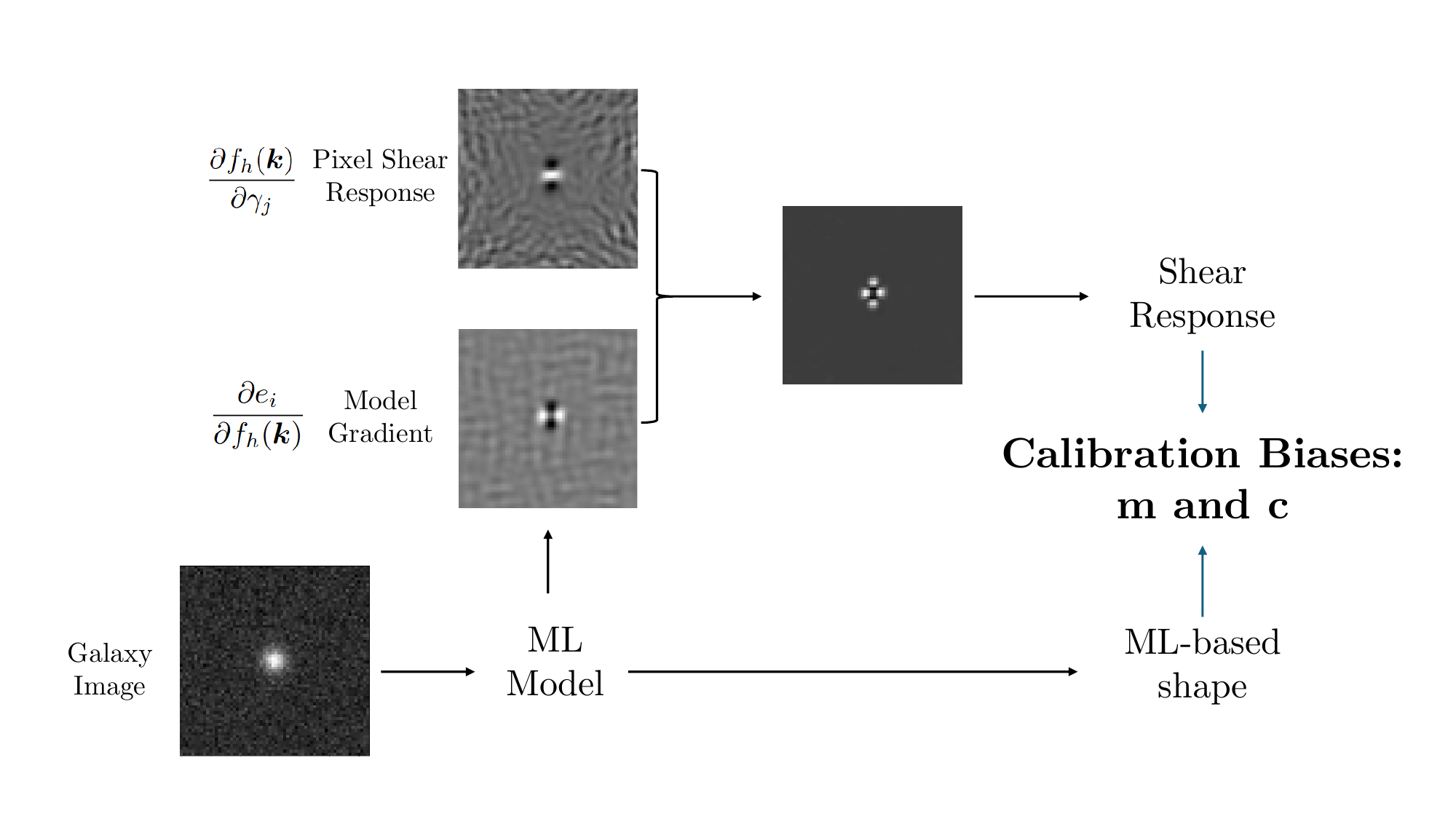}
\caption{
    Workflow of the analytical shear calibration for an ML–based shear
    estimator. Starting from a galaxy image, the ML model produces a raw shape
    estimate and corresponding model gradient via backpropagation. In parallel,
    the shear response of the resmoothed image is computed analytically at the
    pixel level using the expressions in \autoref{eq:pixel_rsp}. 
    The contraction of the model gradient with the pixel shear response yields the
    shear response matrix $R_{ij} = \partial e_i / \partial \gamma_j$, which is
    then used to perform a linear calibration of the ML-based shape estimator.
    The calibrated shear is finally used to evaluate the residual calibration
    biases, quantified by the multiplicative and additive bias parameters $m$
    and $c$.
}
\label{fig:workflow}
\end{figure*}

Based on results from \autoref{subsec:shear_rsp}, for a \Dfour{} symmetric model, the nonlinear bias would be third order after a linear calibration.
In this section, we will introduce an analytical calibration method for a general machine learning model trained with the backpropagation algorithm \citep{Rumelhart1986}.
The key of linear calibration is to calculate the response matrix $R_{ij}$ given by the first-order shear response:
\begin{align}
    R_{ij} &= \pard{e_i}{\gamma_j}  \notag \\
        &= \pard{e_i}{f_h(\vx)} \pard{f_h(\vx)}{\gamma_j}.
\label{eq:rsp_matrix}
\end{align}
The first term on the second row of \autoref{eq:rsp_matrix} is just the gradient of the shape measurement model given the re-smoothed image, which can be easily obtained by backpropagating the model.
The second term is the shear response of the resmoothed image, which can be analytically calculated according to \cite{Li2025}:
\begin{equation}
    \begin{aligned}
    \frac{\partial f_h(\mathbf{x})}{\partial \gamma_1} &= G_1(\mathbf{x}) + x_1 J_1(\mathbf{x}) - x_2 J_2(\mathbf{x}), \\
    \frac{\partial f_h(\mathbf{x})}{\partial \gamma_2} &= G_2(\mathbf{x}) + x_2 J_1(\mathbf{x}) + x_1 J_2(\mathbf{x}),
    \end{aligned}
    \label{eq:pixel_rsp}
\end{equation}
where $G_1, G_2, J_1, J_2$ are linear functions of the image:
\begin{equation}
    \begin{aligned}
    G_1(\mathbf{x}) &= \mathcal{F}^{-1} \left[ \left( \frac{f_p(\mathbf{k})}{p(\mathbf{k})} - \frac{n(\mathbf{k})}{p^{90}(\mathbf{k})} \right) h(\mathbf{k}) \, (k_1^2 - k_2^2) \right], \\[6pt]
    G_2(\mathbf{x}) &= \mathcal{F}^{-1} \left[ \left( \frac{f_p(\mathbf{k})}{p(\mathbf{k})} - \frac{n(\mathbf{k})}{p^{90}(\mathbf{k})} \right) h(\mathbf{k}) \, (2k_1 k_2) \right], \\[6pt]
    J_1(\mathbf{x}) &= \mathrm{i} \, \mathcal{F}^{-1} \left[ \left( \frac{f_p(\mathbf{k})}{p(\mathbf{k})} - \frac{n(\mathbf{k})}{p^{90}(\mathbf{k})} \right) h(\mathbf{k}) \, k_1 \right], \\[6pt]
    J_2(\mathbf{x}) &= \mathrm{i} \, \mathcal{F}^{-1} \left[ \left( \frac{f_p(\mathbf{k})}{p(\mathbf{k})} - \frac{n(\mathbf{k})}{p^{90}(\mathbf{k})} \right) h(\mathbf{k}) \, k_2 \right].
    \end{aligned}
\end{equation}

It's worth noticing that the shear response analysis of \autoref{eq:shear_r} requires the shape to be continuously differentiable to at least the third order. As a result, the machine learning model needs to be a smooth function of the input image, and any module with a discontinuous derivative (e.g., ReLU, Maxpooling) is not allowed in our model.

Combining the results above, $R_{ij} = \pard{e_{i}}{f_h(\vk)}\pard{f_h(\vk)}{\gamma_j}$ can be calculated combining model gradient and \autoref{eq:pixel_rsp}.
As the off-diagonal terms of the averaged response matrix are usually zero in practice,
we will also use $\langle R_i \rangle$ to denote the diagonal term $\langle R_{ii} \rangle$ for simplicity.
In this case, the calibrated unbiased shear estimator would be:
\begin{align}
    \hat{\gamma}_i  = \langle e_i \rangle/\langle R_i \rangle
\end{align}
In principle, for a \textbf{\Dfour{} symmetric} and \textbf{continuously differentiable} ML model, the bias for this estimator would be third order.
The above process is also concluded in \autoref{fig:workflow}.

\subsubsection{Calibration for Selection Bias}
\label{subsec:sel_bias}
For real observations, the ensemble average mentioned in the previous sections is usually computed on a set of galaxies, selected using certain observed variables $\alpha$ (e.g., magnitude, galaxy size, etc.).
Following \citep{kaiser2000Selection}, we interpret such hard-cut-selection as a Heaviside step selection weight:
$$
\begin{aligned}
    w_\mathrm{sel} =
    \begin{cases}
        1, & \text{if } \alpha > \alpha_0, \\
        0, & \text{if } \alpha \le \alpha_0 .
        \end{cases}
\end{aligned}
$$
In this case, the {\color{red}denominator} in \autoref{eq:unbiased_esti} can be rewritten as:
\begin{align}
    \frac{\partial \langle e_{1,2} \rangle}{\partial \gamma_i}
    = \left\langle{w_\mathrm{sel}\pard{e_{1,2}}{\gamma_i}}\right\rangle
    +\left\langle{e_{1,2}\pard{w_\mathrm{sel}}{\gamma_i}}\right\rangle.
\end{align}
However, the $\pard{w_\mathrm{sel}}{\gamma_i}$ in the second term is usually
hard to get. As an alternative way, following the finite-difference approach of
\citet{Sheldon2017}, we take the variable $\alpha$ and its shear response
$\pard{\alpha}{\gamma_i}$, and use
$\alpha\pm\pard{\alpha}{\gamma_i}\gamma_{0,i}$ to mimic the magnitude
measurement under shear $\pm \gamma_{0,i}$. With such a mock, we can apply the
same cut $\alpha_0$ to get the mocked selection weight $w^{\pm}{i}$ under a
small finite shear $\gamma_{0}$. With the mocked weights, the response could be
approximated by:

\begin{align}
    \frac{\partial \langle e_{1,2} \rangle}{\partial \gamma_i}
    = \left\langle{w_\mathrm{sel}\pard{e_{1,2}}{\gamma_i}}\right\rangle
    +\frac{\langle w^+_i e_{1,2} \rangle-\langle w^-_i e_{1,2} \rangle}{2\gamma_{0,i}}.
\end{align}

Combining with the formulas in \autoref{subsec:anacal}, we can have the eventual response with consideration of the selection effect:
\begin{align}
    \langle R_i \rangle = \left\langle \pard{e_i}{f_h(\vk)} \pard{f_h(\vk)}{\gamma_j} \right\rangle+ \frac{\langle w^+_i e_{1,2} \rangle-\langle w^-_i e_{1,2} \rangle}{2\gamma_{0,i}}.
    \label{eq:response_full_selection}
\end{align}
The first term is computed fully analytically via backpropagation, as described in \autoref{subsec:anacal}. The second term uses a finite-difference approximation for the selection weight derivative $\partial w_{\rm sel}/\partial\gamma_i$, following the approach of \citet{Sheldon2017} and \citet{LiMandelbaum2023}; this requires evaluating the estimator on mock-sheared selection variables $\alpha \pm (\partial\alpha/\partial\gamma_i)\gamma_0$ but involves no image re-renderings. The overall framework is therefore analytic at the \textit{measurement} level and semi-analytic at the selection level, requiring no additional image simulations at either stage.

\subsection{Data and Analysis Pipeline}
\label{sec:Data}
In this paper, we focus on testing the performance of our ML-based shear estimator after the calibration process mentioned in \autoref{sec:ML_Anacal}.
To this end, we simulate mock astronomical images distorted by known shear $\gamma_{1,2}$, and compare that to the estimated shear $\hat{\gamma}_{1,2}$.
To quantify the accuracy of the shear estimator, we use the multiplicative bias ($m_{1,2}$) and additive bias ($c_{1,2}$), which are related to the shears as:
\begin{align}
    \hat{\gamma}_{1,2} = (1+m_{1,2})\gamma_{1,2}+c_{1,2}.
\end{align}

As for the precision of the shear estimator, we consider the shape noise:
\begin{align}
    \mathrm{Std}(\hat{\gamma}_i) = \frac{\mathrm{Std}(e_i)}{\langle R_{i} \rangle}.
\label{eq:shape_noise}
\end{align}

We adopt two different definitions of ellipticity as ground-truth labels in the supervised training process.
The first kind is the classical distortion \citep{Bernstein2002shapes}:
\begin{equation}
    e = \frac{1-q^2}{1+q^2},
\end{equation}
where $q=b/a$ is the ratio between the minor axis and the major axis of the disk.
For galaxies with both disk and bulge, we take the luminosity-averaged ellipticity of the two components.
This shape can be directly calculated with the galaxy catalog used to generate the simulation.
We use this definition to train our fiducial model and verify the accuracy of our ML-based estimator as shown in \autoref{sec:Result}.
Another definition comes from Fourier Power Function Shapelets (FPFS) \citep{FPFS2018,FPFS2022}.
We use this definition to benchmark the precision of our model by comparing with classical moment-based results in \autoref{sec:shape_noise}.

In this section, we first describe the setup of our image simulations and then explain our bias and shape noise estimation pipeline in detail.

\subsubsection{Simulation for Isolated Galaxies}


{
\begin{table}
\centering
\caption{Main parameters of the image simulations.}
\label{tab:simulation_parameters_compact}
\renewcommand{\arraystretch}{1.15}
\begin{tabular}{lll}
\toprule
Parameter & Value & Notes \\
\midrule
Input catalog
& DC1 galaxy catalog
&  \\

Pixel scale
& $0.2\,\mathrm{arcsec\,pixel^{-1}}$
& LSST-like sampling \\

Zero Point
& $30$ magnitude
&  \\

Stamp size
& $64\times64$ pixels
& $12.8\times12.8\,\mathrm{arcsec}^{2}$ \\

PSF profile
& Moffat
&\\

Subpixel Offset
& $(-0.1\ \mathrm{arcsec}, 0.1\ \mathrm{arcsec})$
& One pixel length\\

PSF FWHM
& $0.8\,\mathrm{arcsec}$
& Approximately $4$ pixels \\

Calibration scene
& Isolated galaxies
& No blending \\

\bottomrule
\end{tabular}
\label{table:sim_setting}
\end{table}
}

We generate image simulations for isolated galaxies (with no blending effect) using the \xlens{} package, following the settings of \citep{Sheldon2023}, to mimic LSST 10-year coadded images.
\xlens{} \citep{Li2025} uses GalSim \citep{Rowe2015Galsim} to generate realistic, calibrated exposure images consistent with the LSST Science Pipelines (key settings are summarized in \autoref{table:sim_setting}).
Starting from an input Data Challenge 1 (DC1) catalog of galaxies \citep{Sanchez2020}, it renders multi-epoch, multi-band images with configurable WCS, PSF, noise, artifacts, applied shear, and object layouts, providing a modular validation framework for shear-estimation workflows that follows LSST Science Pipelines design patterns.
The noise in each pixel is drawn independently from a Gaussian distribution.
After applying random rotations, shear distortions, subpixel offsets, and PSF convolution, each galaxy is placed within the one pixel square around the center of the $64 \times 64$ pixel postage stamp.
Each simulated image contains $10 \times 10$ postage stamps.
Galaxies are distorted by $(\gamma_1,\gamma_2) = (\pm0.02,0)$ or $(0,\pm0.02)$, meaning there are four different sheared versions for the same galaxy.
To meet the purpose of various tests, we might also add ellipticity to the PSF and random Gaussian noise to the image under specific cases.

For the shear estimation test in \autoref{sec:Result}, we apply a hard cut of $m_\mathrm{i}<25.3$ on input galaxy, consistent with the photometric redshift requirements of LSST \citep{LSSTSRD}, allowing us to focus on a more realistic observational sample. To obtain accurate shear estimation biases, we generate $400{,}000$ images for each sheared configuration, corresponding to $40$ million galaxies per setup.

For the shape noise test in \autoref{sec:shape_noise}, we restrict our analysis to brighter galaxies with $m_\mathrm{i}<24.5$.
To ensure that the shape noise measurement is accurate at the $1\%$ level, we generate $10{,}000$ images, corresponding to $1$ million galaxies in total, for this test.

\subsubsection{Shape noise and image noise cancellation}
To give a tighter measurement of shear estimation biases using fewer simulations, we generate four sheared versions ($(\gamma_1,\gamma_2) = (\pm0.02,0)$ or $(0,\pm0.02)$)
for each galaxy with exactly the same realization of image noise (following \citep{Pujol2019,Sheldon2020}).
At the same time, we generated the galaxies in orthogonal pairs following \cite{Massey2007},
where two galaxies have the same morphology and brightness, but the major axes differ by 90 degrees.
The $40$ million galaxies consist of $20$ million orthogonal pairs.

In this setup, the shear estimation biases can be measured as follow:
\begin{align}
    m_{1,2} = \frac{\langle e^+_{1,2} \rangle-\langle e^-_{1,2} \rangle}
            {0.02(\langle R^+_{1,2} \rangle + \langle R^-_{1,2} \rangle)} -1,
\label{eq:m}
\end{align}
and
\begin{align}
    c_{1,2} = \frac{\langle e^+_{1,2} \rangle+\langle e^-_{1,2} \rangle}
            {\langle R^+_{1,2} \rangle + \langle R^-_{1,2} \rangle},
\label{eq:c}
\end{align}
where $e^{+}_{1,2}$ and $R^+_{1,2}$ are the ellipticity and shear response for the positively shear versions of the galaxies ($(\gamma_1,\gamma_2) = (+0.02,0)$ or $(0,+0.02)$).
Correspondingly, $e^{-}_{1,2}$ and $R^-_{1,2}$ are from the negatively sheared case.
The errors on the biases are estimated with a 20-times bootstrap.

\subsection{Machine Learning Implementation}
\label{sec:ML}

\subsubsection{Model Architecture}
\label{subsec:architecture}

\begin{figure*}[tb]
\centering
\includegraphics[width=1.0\textwidth]{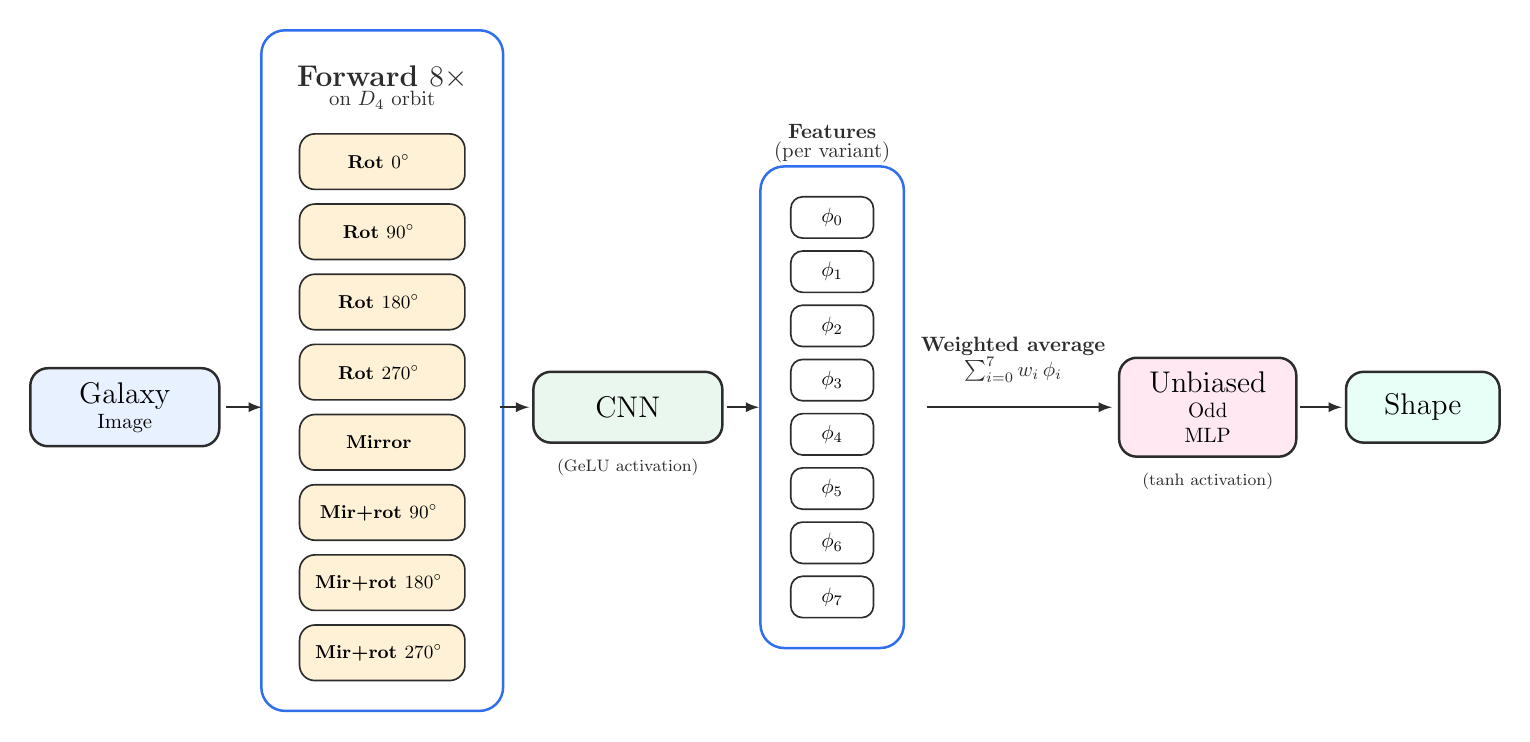}
\caption{
    Example architecture of the \Dfour{}-equivariant shape measurement model (\Dfour{}CNN). The
    input galaxy image is transformed over the full \Dfour{} orbit (four rotations
    and their mirrored counterparts), and we forward the CNN eight times on
    each of the eight transformed variants to produce a set of features. These
    features are then mapped back to the original reference frame and combined
    through a weighted average to construct a \Dfour{}-equivariant feature
    representation. Finally, an unbiased odd MLP preserves the sign of this
    equivariant feature, ensuring the output shape transforms with the desired
    equivariance under \Dfour{} symmetry operations.
}
\label{fig:f8CNN}
\end{figure*}

\begin{figure}[tb]
\centering
\includegraphics[width=0.5\textwidth]{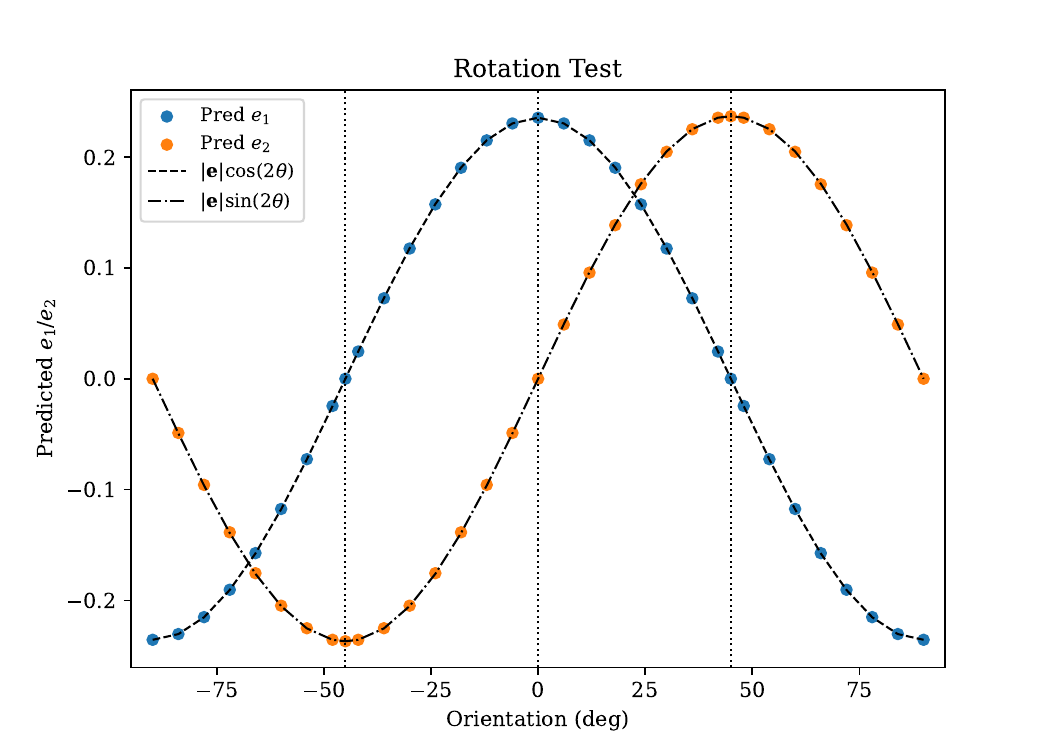}
    \caption{
    Transformation of measured shape under rotation for our \Dfour{}CNN.
    The same galaxy profile with fixed ellipticity ($e = 0.4$) but different orientations ($-90\degree$ to $+90\degree$ between the x-axis and the major axis) is fed into the model.
    The predicted $e_1$ and $e_2$ components (blue and orange points) are compared against theoretical sine/cosine functions (dashed and dash-dotted curves) scaled to match the measured amplitude.
    Vertical dotted lines mark key reference angles at $-45\degree$, $0\degree$, and $+45\degree$.
    }
\label{fig:rot_test}
\end{figure}

As introduced in \autoref{sec:ML_Anacal}, the ellipticity of a pixelized image exhibits rotational and mirror equivariance (i.e.\ \Dfour{} equivariance), which is crucial for unbiased shear estimation.
At the same time, shear response analysis requires the ML model to be continuously differentiable with respect to the input image.
Based on the principles introduced in \autoref{sec:ML_Anacal}, we present an example of a \Dfour-equivariant model in this section to demonstrate the effectiveness of machine-learning-based methods for unbiased shear estimation.

The architecture is shown in \autoref{fig:f8CNN}.
We denote a $90^\circ$ rotation by $\hat{R}_{90}$ and an $x$-axis mirror operation by $\hat{P}$.
The \Dfour{} group then consists of eight elements:
\begin{align}
    \psi_{0-7}  = 1,\hat{R}_{90},\hat{R}^2_{90},\hat{R}^3_{90},
    \hat{P},\hat{P}\hat{R}_{90},\hat{P}\hat{R}^2_{90},\hat{P}\hat{R}^3_{90}.
\end{align}
For any input image $f_h$, the \Dfour{} symmetry generates eight distinct transformed versions, commonly referred to as the orbit, denoted as $\psi_i f_h$ ($i=1,\dots,8$).
Since galaxy shape measurements at high signal-to-noise ratio should be largely insensitive to overall brightness, we normalize the pixel values for bright galaxies to reduce flux-dependent effects.
Because ellipticity information is encoded in the local shape of galaxies without the overall morphology, we adopt a Convolutional Neural Network (CNN) as the backbone architecture.
To construct equivariant features, the same CNN backbone is applied to the whole orbit, resulting in eight forward passes and a corresponding set of feature outputs.
To enforce equivariance, each feature is mapped back to the reference orientation using the inverse transformation, and the final representation is obtained by taking a weighted average over the transformed features.

For a CNN model $F$ that produces feature outputs $\phi_i = F(\psi_if_h)$, we multiply the corresponding weight $w_i = \pm1$ according to the symmetry, and the equivariant features are constructed as
\begin{equation}
\begin{aligned}
\Psi_1
= \frac{1}{8}\big(&
\psi_0^{-1}\phi_0 - \psi_1^{-1}\phi_1 + \psi_2^{-1}\phi_2 - \psi_3^{-1}\phi_3 \\
&+\psi_4^{-1}\phi_4 - \psi_5^{-1}\phi_5 + \psi_6^{-1}\phi_6 - \psi_7^{-1}\phi_7
\big),
\end{aligned}
\end{equation}

\begin{equation}
\begin{aligned}
\Psi_2
= \frac{1}{8}\big(&
\psi_0^{-1}\phi_0 - \psi_1^{-1}\phi_1 + \psi_2^{-1}\phi_2 - \psi_3^{-1}\phi_3 \\
&- \psi_4^{-1}\phi_4 + \psi_5^{-1}\phi_5 - \psi_6^{-1}\phi_6 + \psi_7^{-1}\phi_7
\big),
\end{aligned}
\end{equation}
where $\Psi_1$ and $\Psi_2$ correspond to the equivariant feature components associated with the shape components $e_1$ and $e_2$, respectively.
The equivariant feature maps are first multiplied by a Gaussian kernel and then processed using global average pooling to produce a single scalar feature for each CNN channel.
These equivariant features are subsequently fed into two unbiased odd multilayer perceptrons (MLPs) with \texttt{tanh} activation functions. This design ensures that the final shape measurements are odd functions of the input features, preserving the sign of the ellipticity components and thereby maintaining the desired \Dfour{} equivariance.
In addition, layer-normalization modules \citep{LeiBa2016} are inserted between CNN layers to stabilize the training and improve convergence.

To demonstrate the effect of \Dfour{}-equivariance, in \autoref{fig:rot_test}, we show the rotation test for the measured ellipticity with our \Dfour{} equivariance architecture.
We generate a series of noiseless galaxy postage stamps using the \texttt{Galsim} package with the same light profile but different orientations.
We input those images into the model and compared the output shapes with respect to the expected values from spin-2 symmetry.
As shown, the prediction matches theory expectations.
In particular, the results at $0\degree$ and $\pm90\degree$ agree exactly due to the hard-coded symmetry.
Though we only hard-coded symmetry under $90\degree$-rotation and mirror, the agreement between measurements and theoretical expectations validates the model's transformation property under any rotation.

\subsubsection{Training Strategy}
\label{subsec:training}

\begin{table}[t]
\centering
\caption{Information for hyperparameters and datasets}
\begin{tabular}{lll}
\toprule
\textbf{Category} & \textbf{Hyperparameter} & \textbf{Value} \\
\midrule
\multicolumn{3}{l}{\textit{Model}} \\
 & Number of layers & 5 \\
 & Base channels & 32 \\
 & Residual factor & 0.1 \\
 & Kernel size & 3$\times$3 \\

\multicolumn{3}{l}{\textit{Optimization}} \\
 & Optimizer & AdamW \\
 & Learning rate & $10^{-3}$ \\
 & Weight decay & $10^{-4}$ \\
 & Scheduler factor & 0.316\\
 & Scheduler patience & 5 \\

\multicolumn{3}{l}{\textit{Huber Loss}} \\
 & Huber $\delta$ & $10^{-3}$ \\

\multicolumn{3}{l}{\textit{Training}} \\
 & Training set & 10,000 \\
 & Validation set & 1,000 \\
 & Batch size & 500 \\
 & Epochs & 50 \\
 & Early Stop Threshold & LR$<10^{-5}$ \\
 & Validation frequency & every 10 steps \\

\bottomrule
\end{tabular}
\label{table:hyperparams}
\end{table}

The equivariant models are trained to output the shape measurement of the input galaxy postage stamp.
To ensure that the model is robust across a wide range of signal-to-noise ratios (SNRs),
we randomly sample the noise level of each postage stamp such that the resulting SNR distribution is approximately uniform over the range $[5,40]$.
The simulated postage stamps are then converted into resmoothed postage stamps following the procedure outlined in \autoref{subsec:anacal},
using a Gaussian kernel with a full width at half maximum (FWHM) of $0.85$ arcsec for reconvolution.
We eventually use $10,000$ resmoothed postage stamps together with the ground-truth ellipticity from the simulation as the training set,
and $1,000$ as the validation set.
To verify that the model is not overfit to the small training set and that our results are not sensitive to this choice, we trained additional models with 1,000, 5,000, and 20,000 postage stamps and evaluated calibration biases and shape noise on the same test set (\autoref{app:training_set}). 
We find that calibration biases are consistent with zero at all training-set sizes, though the central values show a mild trend toward smaller $|m|$ with increasing data, with shape noise improving by less than 1\% relative when doubling the training-set size beyond 5,000.
This rapid convergence is a direct consequence of the \Dfour{}-equivariant architecture: by construction, the network is forbidden from learning any non-symmetric feature, reducing the effective number of free parameters by a factor of eight relative to an unconstrained CNN of the same depth. 
The smooth GeLU activations \citep{Hendrycks2016} further constrain the functional form of the shear response, leaving fewer degrees of freedom to be determined from data. 
As a result, the model generalizes efficiently from a training set that would be far too small for a conventional architecture. 
A table showing calibration bias and shape noise for different training set size (1k, 5k, 10k, 20k postage stamps) is provided in \autoref{app:training_set}.

Each model is trained for 50 epochs using the AdamW optimizer \citep{AdamW} and Huber loss \citep{Huber_Loss}.
To balance predictive performance with computational efficiency, particularly with a view toward training and inference on CPU-based clusters, we adopt a compact five-layer CNN with 32 feature channels and approximately $\sim110$k trainable parameters. 
 We explore several candidate values for the learning rate and weight decay. An initial learning rate of $10^{-3}$ and a weight decay of $10^{-4}$ provide a stable convergence and good validation performance, and are consequently adopted throughout this work. 
We adopt a learning rate scheduler that reduces the learning rate by a factor of $0.316$ if the validation loss improves by less than $1\%$ over 5 epochs,
allowing more efficient convergence.
Early stop is adopted once the learning rate falls below $10^{-5}$ to avoid unnecessary computation.
To further increase the diversity of the training data, data augmentation is applied during training,
including random $90^\circ$ rotations, mirroring, and additional noise realizations.
Basic settings of the training set and various hyperparameters are summarized in \autoref{table:hyperparams}.

\section{Results}

\subsection{Accuracy of ML-based Estimator}
\label{sec:Result}

\begin{figure*}[tb]
\centering
\includegraphics[width=0.8\textwidth]{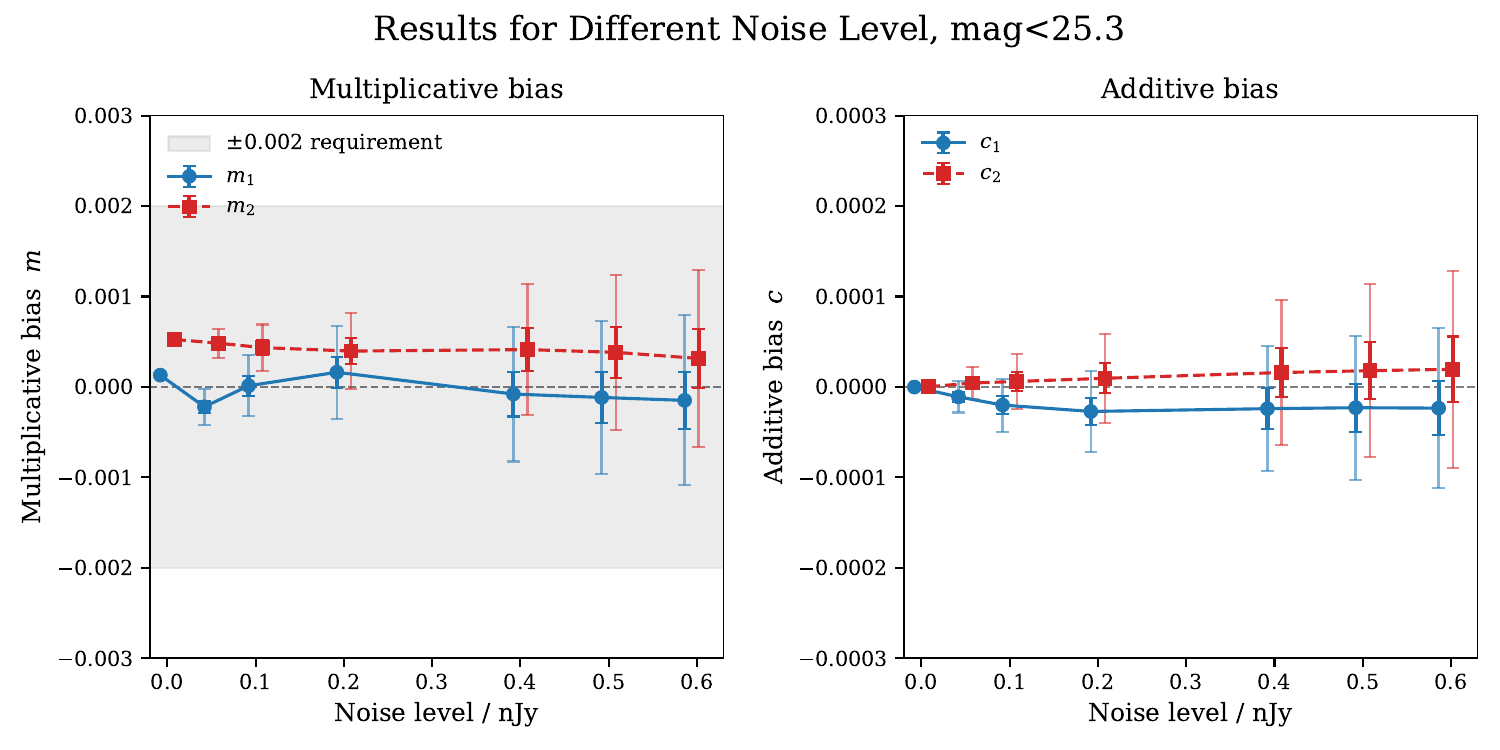}
\includegraphics[width=0.8\textwidth]{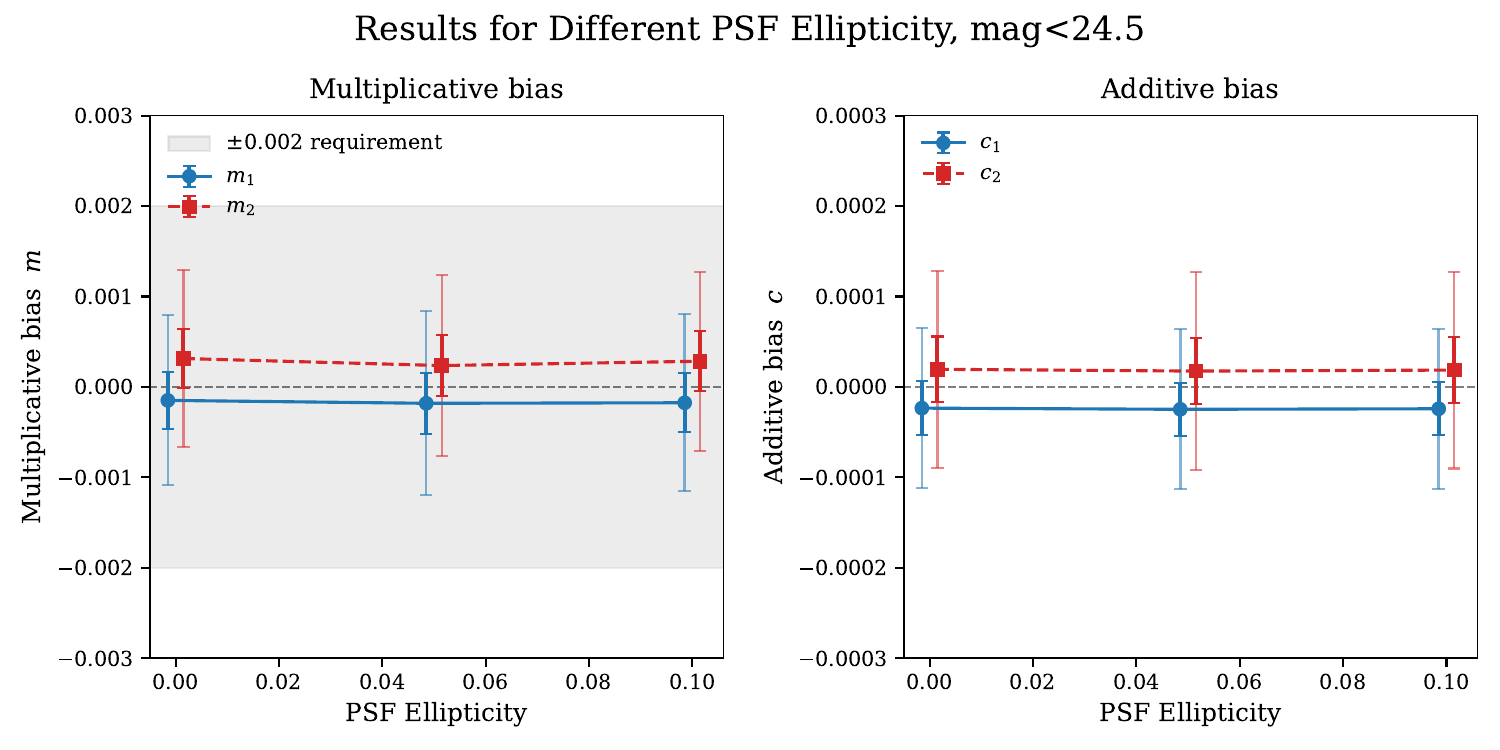}
\includegraphics[width=0.8\textwidth]{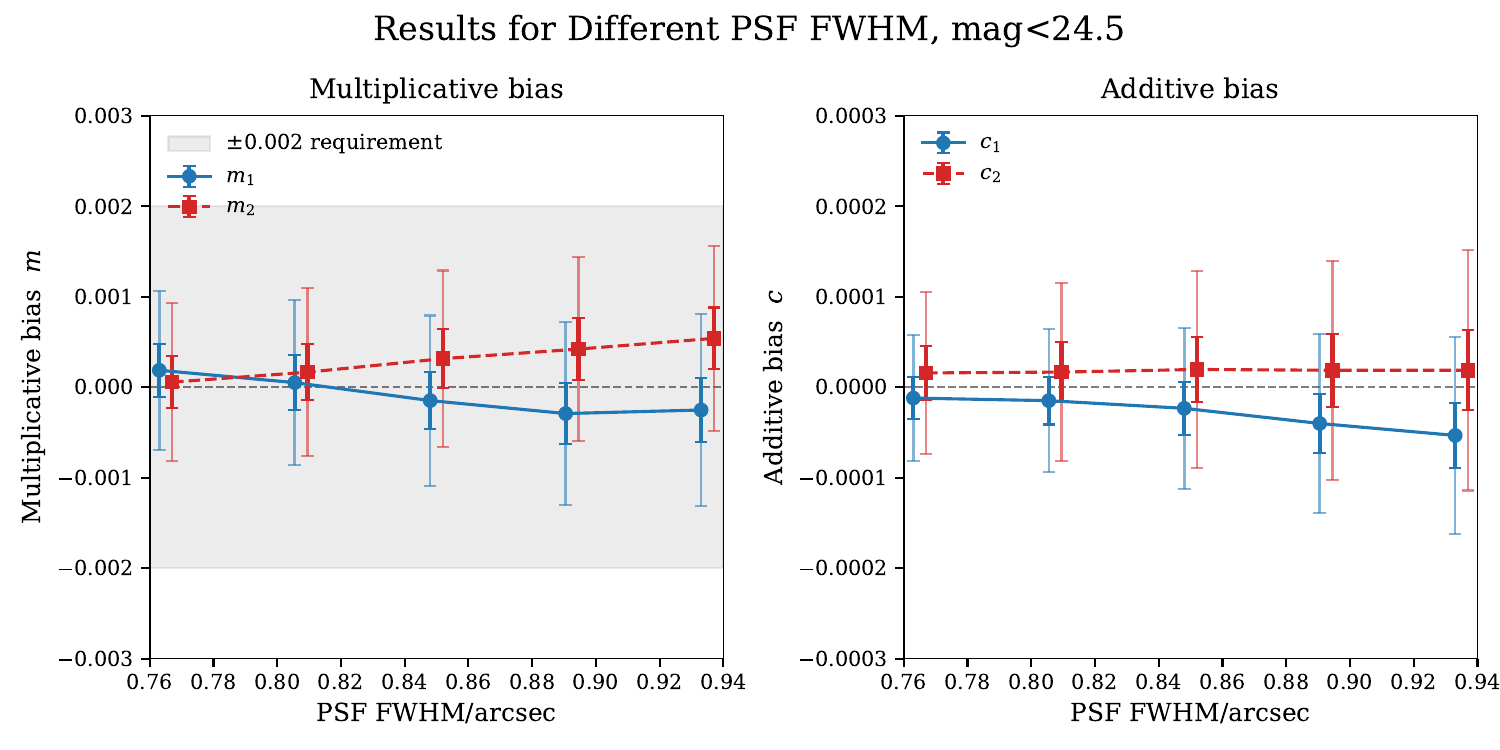}
\caption{
    Shear estimation results for galaxies with $m_i<25.3$ under variations in
    image noise level (top row), PSF ellipticity (middle row), and PSF full
    width at half maximum (bottom row). 
    The left panels show the multiplicative
    bias $m_1$ (blue) and $m_2$ (red), while the right panels show the additive
    bias $c_1$ (blue) and $c_2$ (red). 
    Data points are horizontally slightly
    shifted from the original value to avoid overlap. Points indicate the mean
    bias estimates, with error bars corresponding to $1\sigma$ (deep color) and
    $3\sigma$ (light color) statistical uncertainties with bootstrap resampling
    20 times. The shaded gray region in the multiplicative-bias panels denotes
    the LSST requirement of $|m|<0.002$. Across all tested observing
    conditions, both multiplicative and additive biases remain consistent with
    zero within errors, demonstrating the accuracy of the calibrated ML-based
    shear estimator.
}
\label{fig:m&c}
\end{figure*}

In this section, we focus on the accuracy of the ML-based shear estimator.
We test shear estimation biases of the model under different noise levels, different PSF ellipticities, and PSF sizes.
We first test the effectiveness of calibration for galaxies with $m_\mathrm{i}<25.3$ under the above scenarios,
and then we apply the flux-based selection to test the correction for selection biases.
As a reference, we also report the calibration biases obtained by applying AnaCal to the traditional FPFS shape estimator under identical simulation conditions, which serves as the primary baseline for our D$_4$CNN-based approach.

\subsubsection{Noises and PSF}
\label{subsec:biases}

We summarize the shear estimation bias results in \autoref{fig:m&c}.
These tests are performed using a model trained on $10{,}000$ galaxies, following the training procedure described in \autoref{subsec:training}.
For each data point in the figure, we generate $20$ million paired galaxies, as described in \autoref{sec:Data}, to test the calibration biases.
The left panels show the multiplicative bias components $m_{1,2}$, while the right panels display the additive bias components $c_{1,2}$.

In the top row, we present the shear estimation biases for varying image noise levels, where the largest value, $0.594$ nJy, corresponds to the noise level of LSST Year-10 coadded \textit{i}-band images.
The middle row shows the impact of PSF ellipticity, introduced as $e_1 = e_\mathrm{PSF}$ and $e_2 = -e_\mathrm{PSF}$ following \cite{LiMandelbaum2023}.
In the bottom row, we examine the dependence on PSF size, parameterized by the full width at half maximum (FWHM) of the reconvoluted Gaussian kernel used in the resmoothing process described in \autoref{subsec:anacal}.
The model is trained with a PSF FWHM of $0.85\,\mathrm{arcsec}$, corresponding to the central value in the plots.
We vary the PSF size by $\pm 10\%$ around this fiducial value and evaluate the resulting shear estimation biases.

As shown in \autoref{fig:m&c}, the multiplicative biases in all cases satisfy the LSST requirement of $|m| < 0.002$, indicated by the shaded gray regions.
All measured biases are consistent with zero within the statistical uncertainties, demonstrating that our calibrated ML-based shear estimator is unbiased for isolated galaxies across a range of observing conditions.

\subsubsection{Magnitude Selection}
\label{subsec:result_sel}

\begin{figure*}[tb]
\centering
\includegraphics[width=0.8\textwidth]{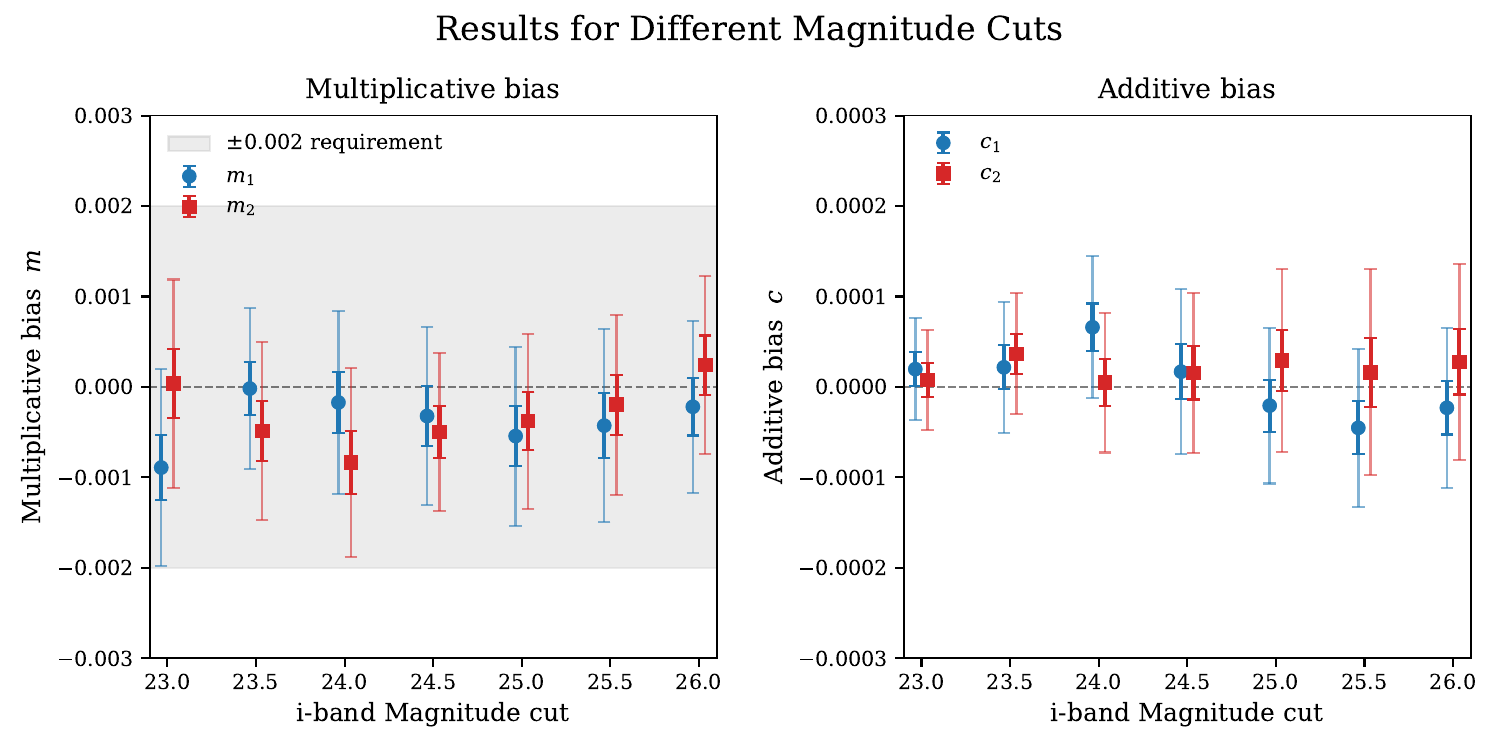}
\caption{
    Shear estimation results for galaxies with different i-band magnitude cuts.
    The shaded gray region in the multiplicative-bias panels denotes the LSST
    requirement of $|m|<0.002$. Left panels show the multiplicative bias $m_1$
    (blue) and $m_2$ (red), while the right panels show the additive bias $c_1$
    (blue) and $c_2$ (red). Points indicate the mean bias estimates, with error
    bars corresponding to $1\sigma$ (deep color) and $3\sigma$ (light color)
    statistical uncertainties with bootstrap resampling 20 times. Data points
    are horizontally slightly shifted from the original value to avoid overlap.
    With different magnitude cuts, both multiplicative and additive biases
    remain consistent with zero within errors, demonstrating the accuracy of
    the calibrated ML-based shear estimator.
}
\label{fig:mag_cuts}
\end{figure*}

Selection biases are crucial in real observations.
To test the selection bias calibration described in \autoref{subsec:sel_bias}, we adopt a hybrid approach: galaxy shapes are estimated by our \Dfour{}CNN model, while the selection variable $\alpha$ (measured flux/magnitude) is taken from the FPFS method \citep{FPFS2018,FPFS2022}. 
This is motivated by the fact that our model outputs ellipticity only and does not independently estimate flux; FPFS provides a well-characterized, AnaCal-compatible flux estimator whose shear response can be analytically computed.
We then apply the finite-difference selection correction, using the FPFS-derived $\alpha$ and its known shear response, with cuts applied at the FPFS-measured magnitude.
This hybrid approach is consistent with how AnaCal is applied in practice and does not require the shape estimator itself to produce flux estimates.
The main result is shown in \autoref{fig:mag_cuts}, where we applied cuts on the measured magnitude from FPFS methods, and include all galaxies satisfying $m_\mathrm{i}<m_\mathrm{cut}$ as samples for the calibration test.
We test magnitude cuts $m_\mathrm{cut} = 23.0, 23.5, 24.0,24.5,25.0,25.5 \!\!\And\!\! 26.0$.
The multiplicative biases in all cases satisfy the LSST requirement of $|m| < 0.002$, indicated by the shaded gray regions.
This result further proves our ML-based shear estimator is unbiased after calibration with selection based on measured magnitude.

\subsection{Precision of ML-based Estimator}
\label{sec:shape_noise}

\begin{figure*}[tb]
\centering
\includegraphics[width=0.8\textwidth]{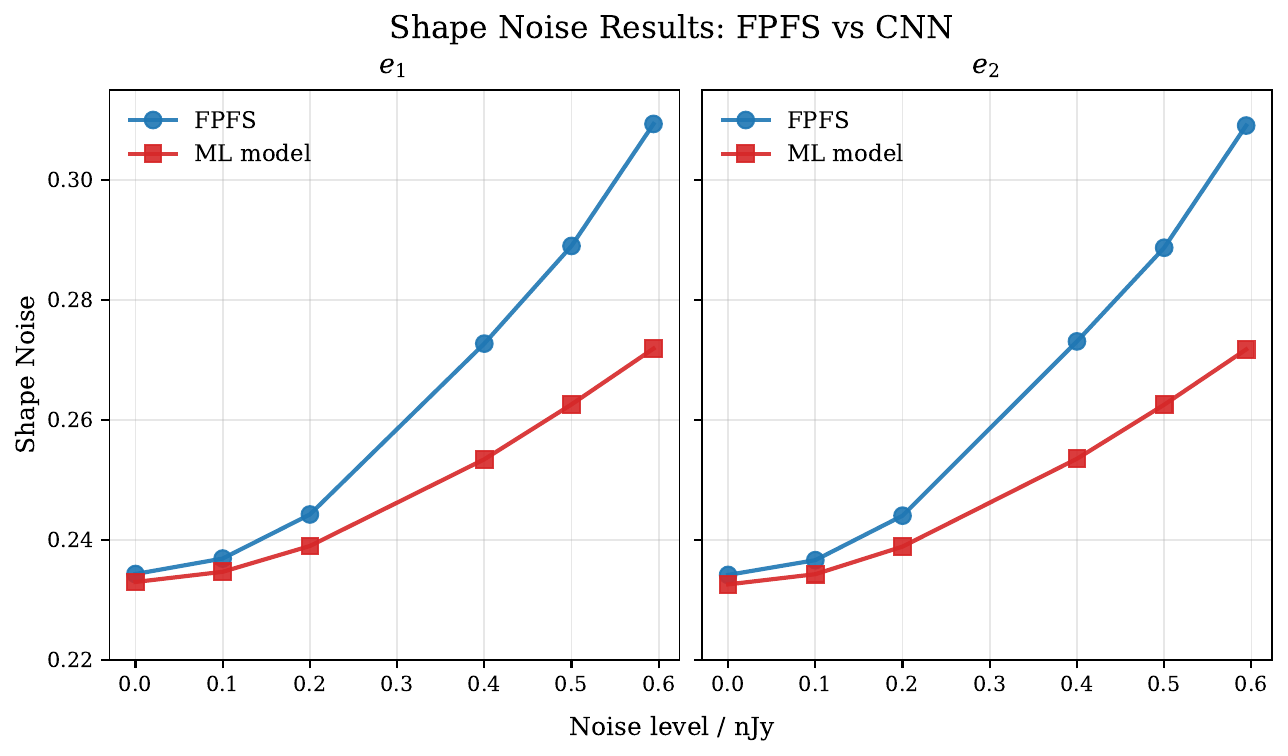}
\caption{
    Shape noise comparison between our CNN model (red squares) and FPFS (blue circles) as a function of image noise level.
    The highest noise level corresponds to the standard deviation expected for LSST 10-year coadded $i$-band images with a magnitude cut $m_\mathrm{cut}=24.5$.
    While both methods achieve comparable shape noise at low noise levels, the CNN model shows a $\sim$10\% reduction in shape noise in the high-noise regime, demonstrating the improved noise robustness and effective denoising capability of the ML-based approach.
}
\label{fig:shape_noise}
\end{figure*}

In this section, we focus on the precision of our ML-based shear estimator.
We use the shape noise introduced in \autoref{eq:shape_noise} to benchmark the precision.
To avoid the statistical precision loss from the different definition of ellipticity,
we train another ML model with the same procedure in \autoref{subsec:training}, but using the FPFS ellipticity on noiseless images as the ground-truth labels.
The shape noise is then measured on $1$ million postage stamps of simulated galaxies.
In \autoref{fig:shape_noise}, we show the results of shape noise of our ML model compared to FPFS method,  with a magnitude cut $m_\mathrm{cut}=24.5$.
The test is done with 1 million galaxies, so the uncertainty of the measured shape noise is at a $\sim10^{-4}$ level.
In the low noise end, our model and FPFS measurement have similar shape noise, as the model is trained with FPFS ellipticity.
But on high-noise images, our model shows lower shape noise and $\sim 10\%$ improvement on LSST 10-year coadded i-band images (the largest noise level in the plots),
which is equivalent to a $23\%$ increase in the galaxy number density.
This indicates that our model adapts to low SNR scenarios better and is able to fetch more shape information from a noisy image compared to the moment-based FPFS method.
For cosmic-shear measurements, the shape-noise power spectrum scales as ($N_\ell\propto \sigma_e^2/n_{\rm eff}$). Thus, a 10\% reduction in the per-component shape dispersion reduces the rms shape noise of the reconstructed shear field by 10\% and the corresponding noise power by approximately 19\%.
In the strictly shape-noise-dominated limit, where the band-power uncertainty scales as $\Delta C_\ell\propto C_\ell+N_\ell\simeq N_\ell$, this would reduce the uncertainty on $C_\ell$ by approximately 19\%; the improvement in final cosmological constraints will generally be smaller once cosmic variance, scale correlations, and systematic uncertainties are included.

\begin{table*}[ht]
\centering
\small
\begin{tabular}{c c c c c c}
\hline
Model & $m_1\,(10^{-3})$ & $m_2\,(10^{-3})$ 
& $c_1\,(10^{-5})$ & $c_2\,(10^{-5})$ 
& shape noise
\\
\hline

D$_4$CNN (w/ GeLU)
& $-0.092 \pm 0.286$
& $0.183 \pm 0.272$
& $-0.17 \pm 2.48$
& $1.96 \pm 2.93$
& $(0.272, 0.272)$
\\

D$_4$CNN (w/ ReLU)
& $-0.224 \pm 0.320$
& $0.176 \pm 0.326$
& $-0.37 \pm 2.64$
& $1.66 \pm 3.24$
& $(0.271, 0.273)$
\\

CNN
& $-0.531 \pm 0.295$
& $-1.388 \pm 0.292$
& $6100.06 \pm 3.59$
& $3700.44 \pm 3.54$ 
& $(0.277,0.275)$
\\

\hline
\end{tabular}
\caption{
Calibration biases and shape noise for the three ML model variants, all evaluated at the LSST Year-10 noise level (0.594\,nJy). D$_4$CNN (w/~GeLU) is the fiducial model trained with FPFS ellipticity; D$_4$CNN (w/~ReLU) and CNN are ablation variants (\autoref{sec:sym&grad}). Uncertainties are standard deviations from 20 bootstrap resamplings.
}

\label{tab:bias_sn}
\end{table*}

\section{Discussion: Importance of \Dfour{} Symmetry and smooth gradient}
\label{sec:sym&grad}

\begin{figure}[tb]
\centering
\includegraphics[width=0.5\textwidth]{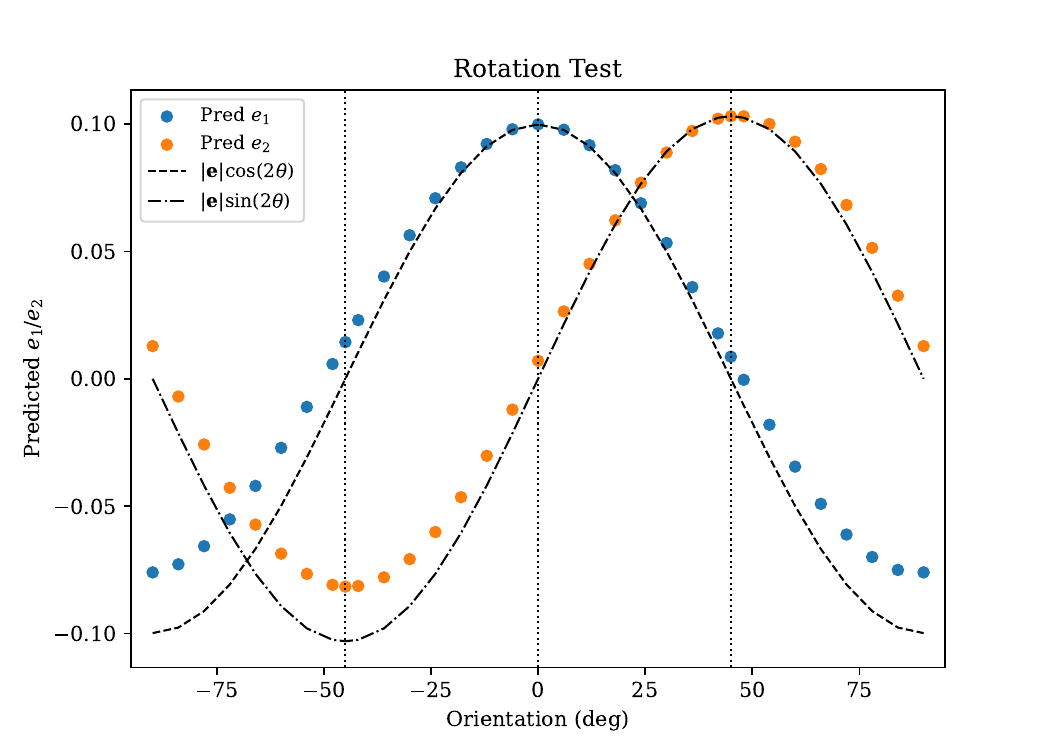}
    \caption{
    Transformation of measured shape under rotation for our CNN model as in \autoref{fig:rot_test}.
    The same galaxy profile with fixed ellipticity ($e = 0.4$) but different orientations ($-90\degree$ to $+90\degree$ between the x-axis and the major axis) is fed into the model.
    The predicted $e_1$ and $e_2$ components (blue and orange points) are compared against theoretical sine/cosine functions (dashed and dash-dotted curves) scaled to match the measured amplitude.
    Vertical dotted lines mark key reference angles at $-45\degree$, $0\degree$, and $+45\degree$.
    }
\label{fig:NonEq_rot_test}
\end{figure}

\begin{figure*}[tb]
\centering
\includegraphics[width=1.0\textwidth]{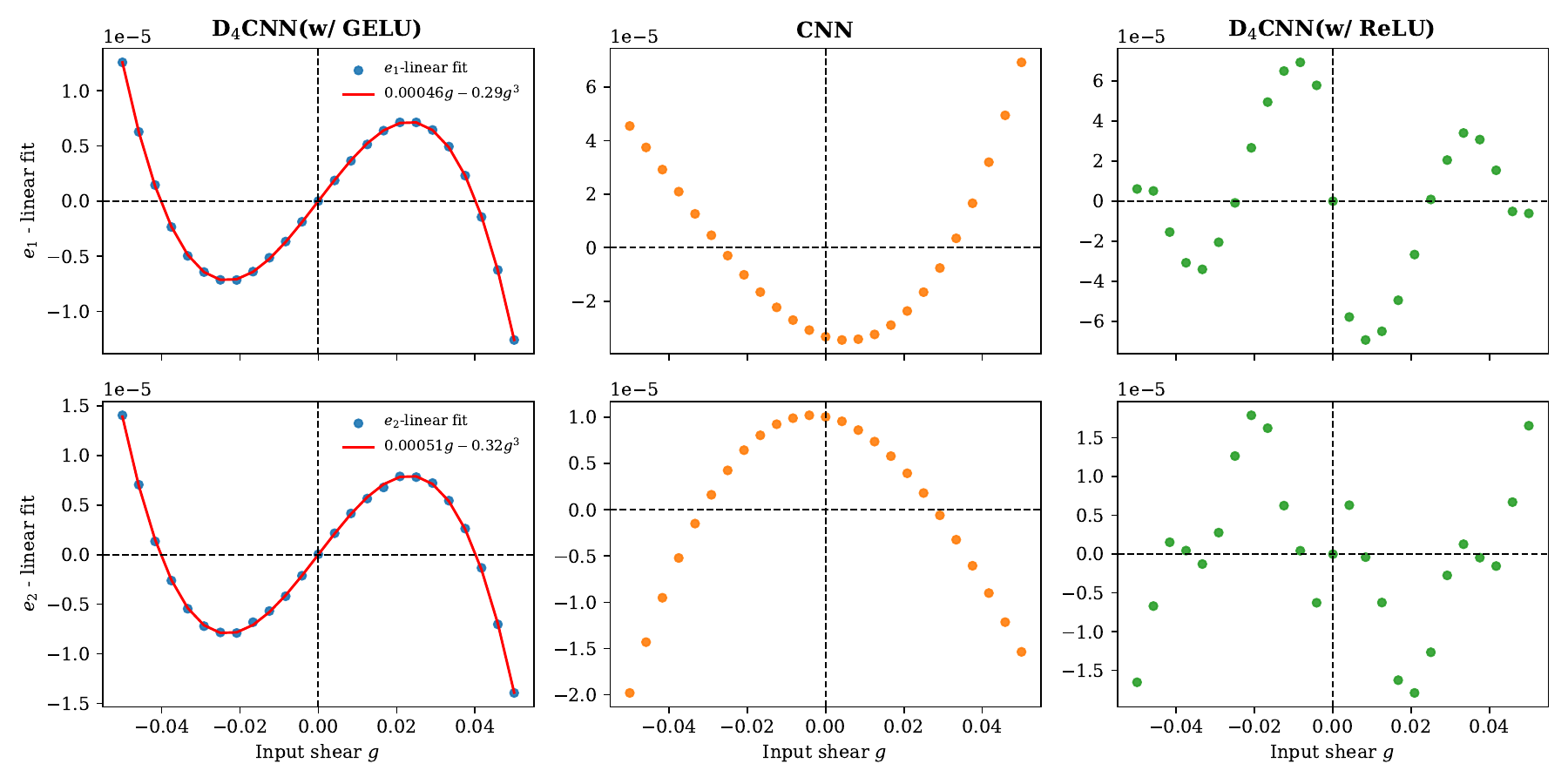}
\caption{
    Non-linear response of the three models to applied shear.
    Galaxies with identical circular profiles but different input shear $g$ are fed into the models. 
    The linear response of the measured ellipticity is fitted and subtracted, and the residual $e_i-\text{linear fit}$ is plotted as a function of the input shear. 
    The top and bottom rows correspond to the two ellipticity components $e_1$ and $e_2$, respectively.
    The \Dfour{}CNN (w/ GELU) (left column) exhibits an approximately cubic behavior with no visible quadratic component (fitted curves are shown in red), consistent with the suppression of even-order terms expected from the imposed symmetry. 
    In contrast, the CNN model (middle column) shows a clear parabolic trend, indicating the presence of second-order nonlinear terms arising from the lack of \Dfour{} equivariance. 
    The \Dfour{}CNN (w/ RELU) (right column), while roughly preserving odd symmetry, displays larger scatter and higher-order nonlinear structures, likely due to the discontinuous gradient of the ReLU activation function. 
}
\label{fig:shearing}
\end{figure*}

In this section, we demonstrate the impact of \Dfour{} symmetry and smooth gradients by comparing the fiducial model described above with models that lack either hard-coded symmetry or smooth activation functions.
For this purpose, we trained two additional models: one with hard-coded symmetry but using the rectified linear unit (ReLU) as the activation function (hereafter referred to as the \Dfour{}CNN (w/ ReLU) model), and another with smooth activation functions but without equivariance (hereafter referred to as the CNN model).
Both models are trained using FPFS ellipticity as the training target to enable a direct comparison with the shape-noise results presented in \autoref{sec:shape_noise}.

The calibration bias and shape-noise results are summarized in \autoref{tab:bias_sn}.
Overall, all three models exhibit similar shape-noise levels. This suggests that the precision improvement discussed in \autoref{sec:shape_noise} primarily originates from the CNN architecture itself rather than from the hard-coded symmetry or smooth activation functions.
However, the CNN model shows a significant additive bias and a slightly larger multiplicative bias. This occurs because the absence of hard-coded symmetry allows the model to develop a zero-point shift in the predicted ellipticity.
This behavior is clearly illustrated by the rotation test for the CNN model shown in \autoref{fig:NonEq_rot_test}.
The plot reveals a noticeable shift in the measured ellipticity, and the measured values deviate from the theoretical prediction due to the lack of \Dfour{} equivariance.

To further compare the nonlinearity of the three models, we feed galaxies with identical circular profiles but different applied shears into each model.
We then perform a linear fit to the measured shapes and subtract the fitted component, allowing the remaining nonlinear response to be examined as a function of shear.

As shown in \autoref{fig:shearing}, 
our smooth and equivariant model exhibits a cubic polynomial behavior with no visible quadratic term. Fitting the residual as $e_i - \text{linear fit} = Ag^3+Bg$, we find $A\approx -0.3 $ for both components, consistent with the expected third-order leading nonlinearity predicted by \autoref{eq:shear_r}. The small residual linear component ($B\approx 5 {\times} 10^{-4}$) is consistent with finite-sample noise in the linear fit and vanishes when averaged over a large ensemble, as confirmed by the near-zero multiplicative biases in \autoref{sec:Result}.
In contrast, the CNN model clearly displays a parabolic trend, indicating the presence of second-order terms.
For the ReLU model, although the response roughly follows the shape of an odd function, the results show larger scatter and noticeable traces of higher-order nonlinear components.

These results are consistent with the theoretical expectations discussed in \autoref{sec:ML_Anacal}. In particular, \Dfour{} equivariance suppresses even-order terms, while smooth activation functions help reduce non-linearity and avoid scattering in the response. 
Moreover, given our testing method in \autoref{eq:m} and \autoref{eq:c}, the multiplicative bias can only contain odd-order non-linearities, whereas all even-order terms contribute instead to the additive bias, as observed for the CNN model. 
Consequently, although the simple CNN architecture used here does not exhibit a significant multiplicative bias, non-symmetric and non-linear effects may accumulate in deeper or more complex models.

It is also worth noting that, for a controlled comparison, we use the same training strategy and architecture for the CNN model. 
Although it is in principle possible to reduce the zero-point shift by changing the architecture or using specific training tricks, the even-order derivatives are difficult to eliminate without explicitly hard-coding the symmetry into the architecture.
Moreover, in our experiments, the CNN model is significantly harder to train. 
Because of the large intrinsic shape noise and the absence of symmetric constraints, the model can easily collapse to a nearly constant output. 
This behavior suggests that hard-coded symmetry may not only reduce systematic biases but also facilitate convergence during training.

\section{Summary and Future Work}
\label{sec:conclusion}

In this paper, we present a physics-informed framework for machine-learning--based shear estimation with the following achievements:

\begin{itemize}

\item \textbf{$\mathbf{D_4}$CNN$\times$\anacal{}.} 
    We develop a general model architecture that explicitly hard-codes the intrinsic \Dfour{} symmetry of shear and calibrates it with the Analytical Calibration (\anacal{}) method, achieving unbiased measurements without requiring image re-renderings at any calibration stage. 
    Measurement-level bias is corrected fully analytically via backpropagation, while selection bias is accounted for using a finite-difference approximation that requires only mock-sheared selection variables rather than additional image simulations (\autoref{subsec:sel_bias}).

\item \textbf{Near-zero bias performance.} 
    Using realistic image simulations of isolated galaxies, we demonstrate that, after calibration, the proposed models achieve near-zero biases across a wide range of noise levels, PSF conditions, and magnitude selection cuts.
    In particular all multiplicative biases satisfying $|m| < 10^{-3}$ and most measurements consistent with zero at the ${\sim}10^{-4}$ level, well within the 0.2\% \textit{LSST} requirement.

\item \textbf{Improved precision.} 
    Our model achieves $\sim$10\% lower shape noise in the high-noise regime, equivalent to a 23\% gain in effective galaxy number density.

\item \textbf{High efficiency.} 
    Our model requires approximately 10 minutes of training on a single NVIDIA H100 GPU and 1-1.5 ms per galaxy for shape measurement and calibration in total, including I/O time.
    For comparison, Metacalibration requires rendering 4--5 sheared image versions per galaxy and re-running the full measurement pipeline on each; at a typical cost of $\sim$5--10\,ms per re-rendering pass \citep{Yamamoto2023}, this amounts to $\sim$25--50\,ms per galaxy, roughly 15--50$\times$ more expensive than our 1-1.5\,ms per galaxy. 
    The FPFS+AnaCal baseline has a comparable per-galaxy cost to our method ($\sim$0.5--1\,ms), but poorer precision as shown in \autoref{sec:shape_noise}.
    
\end{itemize}

Over the past decade, a variety of machine-learning approaches have been proposed for shear estimation.
However, most existing methods do not explicitly enforce the symmetry properties of galaxy shapes, leading to a strong reliance on data-dependent weighting during calibration.
As a result, these methods often struggle to meet the sub-percent accuracy requirements of Stage-IV surveys and may exhibit degraded performance when applied to datasets with different observational conditions \citep{Sheldon2020,DESY6}.
To avoid these issues, we explicitly incorporate the correct symmetry into the model architecture, and build the model with smooth modules only.
Without the loss in precision, this construction strictly controls the calibration biases within the requirement, and potentially helps the training process.
At the same time, the integration of \anacal{} enables a fast and self-consistent calibration procedure for machine-learning models.
Furthermore, the workflow introduced in this work is not limited to the specific CNN architecture presented here. 
More generally, the framework can be applied to a wide range of machine-learning models, provided that the required symmetry is enforced and the model contains only smooth, differentiable modules.

With the promising results for single-band isolated galaxies, future work will extend the framework to blended sources and multi-band observations. These extensions are non-trivial: at LSST depth, blending affects $\sim$50--60\% of detected sources \citep{Dawson2016,SanchezJ2021}, and blend-induced multiplicative bias can reach $|m|{\sim}10^{-2}$ for stamp-based estimators without deblending corrections \citep{Sheldon2020}, i.e., two orders of magnitude larger than the measurement-level biases demonstrated here. Scene-native AnaCal \citep{LiMandelbaum2023} provides an analytic path to propagating blended-neighbor pixels into the shear response matrix, making it the natural next step for the framework. For multi-band data, the \Dfour{}-equivariant architecture extends directly to joint multi-band inputs by treating bands as additional input channels, with the same symmetry constraints applying independently in each band.

For blended sources, the D$_4$CNN will need to be retrained on scene-level inputs with per-source segmentation masks (e.g., from DeepDISC; \citealt{Merz2023,Merz2025}) suppressing cross-source gradient leakage in response, and the finite-difference selection correction of \autoref{eq:response_full_selection} will be extended to a detection-based selection variable. Validation will require demonstrating $|m| {<} 10^{-3}$ on blend-heavy simulations as a function of neighbor flux ratio and separation.

Two additional idealizations warrant future study. First, we assume a perfectly known PSF; in practice, PSF model errors introduce an additive bias $\delta c \sim \alpha_{\rm PSF} \delta e_{\rm PSF}$, where $\alpha_{\rm PSF} \sim 0.01$--$0.05$ is the PSF leakage coefficient and $\delta e_{\rm PSF}$ is the PSF shape residual \citep{Berlfein2025}; validation will require a null test of $|\Delta m_{\rm red-blue}| {<} 2 {\times} 10^{-4}$ between red and blue galaxy subsamples under realistic chromatic PSF models. Second, we assume pure Gaussian pixel noise; correlated noise in coadded images introduces a noise-bias residual that AnaCal corrects analytically via the re-smoothing and re-noising procedure of \autoref{eq:resmooth_img} \citep{Li2025}, but this correction needs to be validated for the non-Gaussian noise distributions (e.g., Poisson shot noise from bright neighbors, cosmic ray residuals) present in real survey data.

Finally, the CNN backbone used here is intentionally minimal (i.e., five layers with base channel width 32) to establish a clean proof of concept. The framework accommodates more powerful backbones: a deeper residual network \citep{HeK2016} is a drop-in replacement that maintains full differentiability, while a vision transformer \citep{Dosovitskiy2020,Paul_Chen_2022} is compatible provided overlapping convolutional patch embeddings and relative positional encodings are adopted to preserve spatial differentiability and \Dfour{} equivariance across the orbit. Either upgrade is expected to reduce shape noise further by extracting higher-order morphological features beyond the reach of moment-based estimators, and the combination of hard-coded physical symmetry, analytic self-calibration, and architectural flexibility positions \textsc{D$_4$CNN}$\times$\textsc{AnaCal} as a practical and principled foundation for machine-learning-based weak lensing shear estimation in the era of Euclid, LSST, and Roman.

\begin{acknowledgments}
We acknowledge Matthew Becker for valuable discussions that motivated the idea of integrating AnaCal with JAX/PyTorch.
We thank Martin Kilbinger, Erin Sheldon, Rachel Mandelbaum, Tianqing Zhang, and Wentao Luo for helpful feedback and comments.
We thank S. Luo and B. Bode at the National Center for Supercomputing Applications
(NCSA) for helpful discussion and assistance with the GPU cluster used in
this work.
Shurui Lin and Xin Liu acknowledge support from the Illinois Campus Research Board
Award RB25035, NSF grant AST-2308174, and NASA grants 80NSSC24K0219 and 80NSSC26K0333.
Xiangchong Li acknowledges support from the U.S. Department of Energy under
Contract No. DE-SC0012704 and from the Laboratory Directed Research and
Development (LDRD) Program at Brookhaven National Laboratory (Project No.
27992).

This work utilizes resources supported by the National Science Foundation's
Major Research Instrumentation program, grant \#1725729, as well as the
University of Illinois at Urbana-Champaign.
This work used Delta and DeltaAI at NCSA through allocations PHY240290 and
PHY250308 from the Advanced Cyberinfrastructure Coordination Ecosystem:
Services \& Support (ACCESS) program, which is supported by U.S. National
Science Foundation grants \#2138259, \#2138286, \#2138307, \#2137603, and
\#2138296.

The code corresponding to this work can be found at \textit{https://github.com/all2b9s/wl\_D4T\_CNN}.
\end{acknowledgments}

\facilities{Rubin LSST}
\software{astropy \citep{2013A&A...558A..33A,2018AJ....156..123A,2022ApJ...935..167A},
}

\appendix

\section{Impact of training set size}
\label{app:training_set}

\begin{table*}[ht]
\centering
\small
\begin{tabular}{c c c c c c}
\hline
Size & $m_1\,(10^{-3})$ & $m_2\,(10^{-3})$ 
& $c_1\,(10^{-5})$ & $c_2\,(10^{-5})$ 
& shape noise
\\
\hline

$1{,}000$
& $-0.537 \pm 0.377$
& $0.423 \pm 0.353$
& $-1.20 \pm 2.76$
& $1.50 \pm 2.95$
& $(0.277, 0.274)$
\\

$5{,}000$
& $-0.472 \pm 0.355$
& $0.310 \pm 0.327$
& $4.84 \pm 2.47$
& $1.99 \pm 3.05$
& $(0.274, 0.274)$
\\

$10{,}000$
& $-0.092 \pm 0.286$
& $0.183 \pm 0.272$
& $-0.17 \pm 2.48$
& $1.96 \pm 2.93$
& $(0.272, 0.272)$
\\

$20{,}000$
& $-0.021 \pm 0.277$
& $0.052 \pm 0.277$
& $0.03 \pm 2.44$
& $1.19 \pm 2.99$
& $(0.271, 0.270)$
\\

\hline
\end{tabular}
\caption{
Calibration biases and shape noise for four \Dfour{}CNN models trained on different datasets, all evaluated at the LSST Year-10 noise level (0.594\,nJy). The training sets are generated using exactly the same procedure, differing only in size. Uncertainties are standard deviations from 20 bootstrap resamplings.
}

\label{table:training_set}
\end{table*}

In this appendix, we present the calibration biases and shape noise for models trained on datasets of varying sizes (1k, 5k, 10k, and 20k postage stamps), as summarized in \autoref{table:training_set}.
Overall, the training set size has only a minor impact on model performance. Even the model trained with just $1{,}000$ galaxies remains consistent with zero bias, highlighting the effectiveness of the built-in \Dfour{} symmetry in enforcing unbiased predictions.
The shape noise shows very limited variation across different training sizes, with a standard deviation of approximately $3 \times 10^{-4}$.
This indicates only a weak dependence of shape noise on the size of the training set.

\bibliography{sample701}{}
\bibliographystyle{aasjournalv7}
\end{document}